# THE EFFECT OF $Z'$ BOSON ON SAME-SIGN DIMUON CHARGE ASYMMETRY IN $B_q^0 - \overline{B_q^0}$ SYSTEM

S. Sahoo[*], M. Kumar and D. Banerjee

[1]Department of Physics, National Institute of Technology,
Durgapur-713209, West Bengal, India
[*]E-mail: sukadevsahoo@yahoo.com

**Abstract**

The recent observation of the same-sign dimuon charge asymmetry in the *B* system by the D0 collaboration has $3.9\sigma$ deviation from the standard model prediction. However, the recent LHCb data on $B_s$ neutral-meson mixing do not accommodate the D0 collaboration result. In this paper, considering the effect of $Z'$-mediated flavour-changing neutral currents in the $B_q^0 - \overline{B_q^0}$ (q = d, s) mixing, the same-sign dimuon charge asymmetry is calculated. We find the same-sign dimuon charge asymmetry is enhanced from its SM prediction and provides signals for new physics beyond the SM.

## 1. Introduction

In the standard model (SM) of particle physics, flavour-changing neutral current (FCNC) processes occur only at the loop-level and are very sensitive to new physics (NP) beyond the SM.[1] The rate of these processes is suppressed by small electroweak gauge coupling, CKM matrix[2,3] elements and loop factors.[4] These suppression factors can be enhanced in NP models. The FCNC processes of *K*, $B_d$ and $B_s$ mesons[5] are still large enough to be studied experimentally as well as theoretically. The $B_q^0 - \overline{B_q^0}$ mixing (q = d, s), meson-antimeson mixing,[6–8] plays an outstanding role in this direction.

The same-sign dimuon charge asymmetry from the semi-leptonic $(s\ell)$ decay of $B_{s,d}$ meson is given by[9,10]:

$$A_{s\ell}^b = \frac{N^{++} - N^{--}}{N^{++} + N^{--}}, \tag{1}$$

where $N^{++}$ corresponds to each B hadron decaying semi-leptonically to $\mu^+ X$ and similarly $N^{--}$ to $\mu^- X$. Recently, the D0 collaboration[11] observed the asymmetry of

$$A_{s\ell}^b = (-7.87 \pm 1.96) \times 10^{-3}. \quad (2)$$

This number is almost 33 times larger than the SM prediction[12],

$$\left(A_{s\ell}^b\right)_{SM} = (-2.4 \pm 0.4) \times 10^{-4}. \quad (3)$$

In order to explain this observed asymmetry, additional CP violation source is strongly required in $B_{s,d}$ mixing. This phenomenon has triggered a lot of investigations in both SM and NP models.

Before the first observation[13] of the D0 charge asymmetry in 2010, the NP models have tried to suppress the additional CP violating or FCNC source[14,15]. After the D0 experiment results, researchers have tried to obtain a sizable NP contribution in different models such as the lepto-quark models[16], the MSSM with non-minimal flavor violation[17–19], R-parity violating supersymmetric model[20], split SUSY model[21], $Z'$ model[20,22] and a fourth generation model[23]. In this paper, we use the $Z'$ model to study the enhancement of the same-sign dimuon charge asymmetry.

$Z'$ bosons are known to exist naturally in well-motivated extensions of the SM.[24-28] Theoretically it is predicted that they exist in grand unified theories (GUTs), left-right symmetric models, Little Higgs models, superstring theories and theories with large extra dimensions. $Z'$ boson is not found so far. The mass of the $Z'$ boson is not known. However, there are stringent limits on the mass of an extra $Z'$ and the $Z-Z'$ mixing angle $\theta$ from the non-observation of direct production at the CDF[29–31] and indirect constraints from the precision data (weak neutral current processes and LEP II).[32–34] These limits are model-dependent, $M_{Z'} \geq 500$ $GeV$ and $|\theta| \leq 10^{-3}$ for standard GUT models and $M_{Z'}$ is of the order of 1 TeV in models with nonuniversal flavor gauge interactions.[35–38]

Several NP theories predict more than one extra neutral gauge bosons and many new fermions. These new (exotic) fermions can mix with the SM fermions and induce FCNCs.[39,40] Mixing between ordinary (doublet) and exotic singlet left-handed quarks induces Z-mediated FCNC. In these models[41–43], one introduces an additional vector-singlet charge −1/3 quark $h$, and allows it to mix with the ordinary down-type quarks d, s and b. Since the weak isospin of the exotic quark is different from that of the ordinary quarks, Z-mediated FCNCs are induced. The Z-mediated FCNC couplings $U_{ds}^Z$, $U_{db}^Z$ and $U_{sb}^Z$ which are in general complex, are constrained by a variety of processes. $U_{ds}^Z$ is bounded by the measurements of $\Delta M_K$ ($K^0$-$\overline{K}^0$ mixing), $|\epsilon_K|$ and $K_L \to \mu^+\mu^-$,[41–43] while constraints on $U_{db}^Z$ and $U_{sb}^Z$ come principally from the experimental limit on B $\left(B \to \ell^+\ell^- X\right)$.[44–47] The constraints on $U_{db}^Z$ and $U_{sb}^Z$ allow significant contributions to $B_q^0 - \overline{B_q^0}$ mixing (q = d, s). NP models with exotic fermions also predict the existence of additional neutral $Z'$ gauge bosons. The mixing among particles which have different $Z'$ quantum numbers will induce

FCNCs due to $Z'$ exchange.[48–50] With FCNCs, the $Z'$ boson contributes at tree level, and its contribution will interfere with the SM contributions. The FCNC has effect in the b-s and b-d sectors,[51] for example the $B_q^0 - \overline{B_q^0}$ mixing and rare decays like $B_q^0 \to \mu^+\mu^-$ (q = d, s). The branching ratio of these rare decays has been measured recently by the LHCb collaboration[52,53] and the CDF collaboration.[54]

In this paper, we analyze the $B_q^0 - \overline{B_q^0}$ mixing in a $Z'$ model to explain the large same-sign dimuon charge asymmetry observed by the D0 experiment. The paper is organised as follows: In Sec. 2, we discuss the phenomenology of $B_q^0 - \overline{B_q^0}$ mixing (q = d, s). In Sec. 3, we discuss about the model we have used with considering contributions from the $Z'$ boson. In Sec. 4, we evaluate the $B_s^0 - \overline{B_s^0}$ and $B_d^0 - \overline{B_d^0}$ mixing mass differences. In Sec. 5, we calculate the same-sign dimuon charge asymmetry in the $B_q^0 - \overline{B_q^0}$ (q = d, s) system considering NP contributions from $M_{12}^q$ and $\Gamma_{12}^q$, and with different masses of $Z'$ boson. Finally, we present our conclusions in Sec. 6.

## 2. The $B_q^0 - \overline{B_q^0}$ Mixing and Same-Sign Dimuon Charge Asymmetry

In the SM, the $B_q^0 - \overline{B_q^0}$ mixing at the lowest order is described by box diagrams involving two W bosons and two up-type quarks (Fig. 1).[6,55] In this case, since the large B mass is off the region of hadronic resonances, the long range interactions arising from intermediate virtual states are negligible. In the SM, $M_{12}$ and $\Gamma_{12}$ are computed from the box diagram and read as:[6,56,57]

$$M_{12}^{SM}(B_q) = \frac{G_F^2 M_W^2 M_{B_q} \eta_{B_q}}{12\pi^2} f_{B_q}^2 B_{B_q} S_0(x_t)\left(V_{tq}^* V_{tb}\right)^2, \qquad (4)$$

$$\Gamma_{12} = \frac{G_F^2 m_b^2 \eta_B' M_{B_q} f_{B_q}^2 B_{B_q}}{8\pi} \times \left[\left(V_{tq}^* V_{tb}\right)^2 + V_{tq}^* V_{tb} V_{cq}^* V_{cb}\, O\!\left(\frac{m_c^2}{m_b^2}\right) + \left(V_{cq}^* V_{cb}\right)^2 O\!\left(\frac{m_c^4}{m_b^4}\right)\right]$$
$$,\ldots\ldots\ldots\ldots\ldots\ldots (5)$$

where $M_{12}$ and $\Gamma_{12}$ are the off-diagonal elements of the mass and decay matrices, $G_F$ is the Fermi constant, $M_W$ is the W boson mass, $m_i$ is the mass of quark i, $x_t = m_t^2/M_W^2$; $M_{B_q}$, $f_{B_q}$ and $B_{B_q}$ are the $B_q^0$ mass, weak decay constant and bag parameter respectively. The Inami–Lim function[58] $S_0(x_t)$ is approximated to be 0.784 $x_t^{0.76}$, $V_{ij}$ are the elements of the CKM matrix;[2,3] $\eta_B$ and $\eta_B'$ are QCD corrections. Both $M_{12}$ and $\Gamma_{12}$ depend on CKM matrix elements.

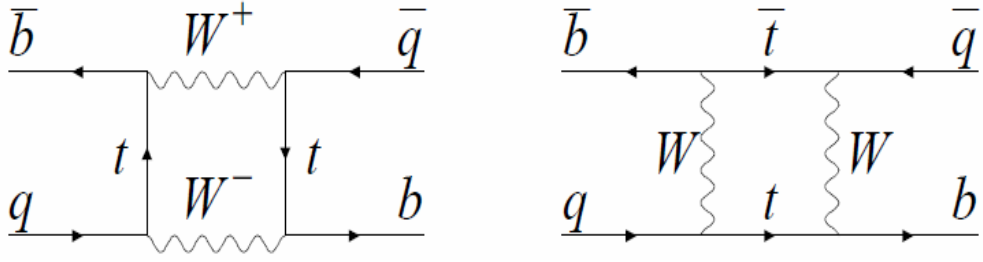

**Fig. 1:** Box diagrams for $B_q^0 - \overline{B_q^0}$ mixing (q = d, s).

The mass difference and decay width difference of the physical 'heavy' (H) and 'light' (L) mass eigenstates are given by the off-diagonal elements by[59]

$$\Delta M_q \equiv M_H(B_q) - M_L(B_q) = 2|M_{12}^q|, \tag{6}$$

$$\Delta \Gamma_q \equiv \Gamma_L(B_q) - \Gamma_H(B_q) = 2|\Gamma_{12}^q|\cos\phi_q. \tag{7}$$

In the SM, the mass difference[60] $\Delta M_{B_d}$ is proportional to the combination of CKM matrix elements $(V_{td}^* V_{tb})^2$. Since the matrix element $V_{ts}$ is larger than $V_{td}$, the expected mass difference $\Delta M_{B_s}$ is higher. In the SM, $B_s^0 - \overline{B_s^0}$ and $B_d^0 - \overline{B_d^0}$ mass differences are found to be:[61]

$$(\Delta M_{B_d})_{SM} = (0.543 \pm 0.091) \text{ ps}^{-1}, \tag{8}$$

$$(\Delta M_{B_s})_{SM} = (17.30 \pm 2.6) \text{ ps}^{-1}, \tag{9}$$

From the recent experiments, $B_s^0 - \overline{B_s^0}$ and $B_d^0 - \overline{B_d^0}$ mass differences are found to be:[61–65]

$$\Delta M_{B_d} = 0.507 \pm 0.004 \text{ ps}^{-1} \text{ (ALEPH, CDF, D0, DELPHI, L3, OPAL,}$$
$$\text{BABAR, BELLE, ARGUS, CLEO),} \tag{10}$$

$$\Delta M_{B_s} = 17.73 \pm 0.05 \text{ ps}^{-1} \text{ (CDF, D0, LHCb),} \tag{11}$$

The CP phase difference between $M_{12}$ and $\Gamma_{12}$ is defined as:

$$\phi_q = \phi_M - \phi_\Gamma = \arg\left(-\frac{M_{12}^q}{\Gamma_{12}^q}\right). \tag{12}$$

In the SM, this angle is found to be:[5]

$$\phi_d^{SM} = \left(-10.1_{-6.3}^{+3.7}\right)\times 10^{-2}\,\text{rad}, \quad \phi_s^{SM} = \left(7.4_{-3.2}^{+0.8}\right)\times 10^{-3}\,\text{rad}. \tag{13}$$

From equations (6) and (7), it is clear that the mass eigenstates have mass and width differences of opposite signs. The heavy state is expected to have smaller decay width than that of the light state. Hence, $\Delta\Gamma = \Gamma_L - \Gamma_H$ is expected to be positive in the SM.[5] Recently, the LHCb collaboration[66] has found that $\Delta\Gamma_s$ is positive.

The wrong-sign charge asymmetries appear in the semileptonic $B_d$ and $B_s$ decays as:[9,10]

$$a_{s\ell}^d \equiv \frac{\Gamma(\bar{B}_d \to \mu^+ X) - \Gamma(B_d \to \mu^- X)}{\Gamma(\bar{B}_d \to \mu^+ X) + \Gamma(B_d \to \mu^- X)}, \tag{14}$$

$$a_{s\ell}^s \equiv \frac{\Gamma(\bar{B}_s \to \mu^+ X) - \Gamma(B_s \to \mu^- X)}{\Gamma(\bar{B}_s \to \mu^+ X) + \Gamma(B_s \to \mu^- X)}. \tag{15}$$

These individual flavor-specific CP asymmetries contribute to the total asymmetry $A_{s\ell}^b$ as:[11,61]

$$A_{s\ell}^b = (0.594 \pm 0.022)\,a_{s\ell}^d + (0.406 \pm 0.022)\,a_{s\ell}^s. \tag{16}$$

In the SM, the individual flavor-specific CP asymmetries are:[61]

$$\left(a_{s\ell}^d\right)_{SM} = -(4.1\pm 0.6)\times 10^{-4}, \quad \left(a_{s\ell}^s\right)_{SM} = -(1.9\pm 0.63)\times 10^{-5}. \tag{17}$$

Hence,

$$\left(A_{s\ell}^b\right)_{SM} = (0.594\pm 0.022)(a_{s\ell}^d)_{SM} + (0.406\pm 0.022)(a_{s\ell}^s)_{SM}$$

$$= (-2.4 \pm 0.4)\times 10^{-4}. \tag{18}$$

From the recent LHCb result[67,68], it is found that

$$a_{s\ell}^s = (-0.24 \pm 0.54\,(\text{stat}) \pm 0.33\,(\text{syst}))\times 10^{-2}. \tag{19}$$

Recently the D0 collaboration[69] has measured the semileptonic charge asymmetry $a_{s\ell}^d$ in $B_d^0$ meson mixing:

$$a_{s\ell}^d = (0.68 \pm 0.45\,(\text{stat}) \pm 0.14\,(\text{syst}))\times 10^{-2}. \tag{20}$$

From equations (2) and (18), it is clear that the experimental value of $A_{s\ell}^b$ is different from the SM value. In order to solve this discrepancy, precise measurements of $a_{s\ell}^s$ and $a_{s\ell}^d$ are needed. The recent study of the CP asymmetry in the $B_s \to J/\psi\phi$ decays[70,71] at the LHCb severely constrains the value of $A_{s\ell}^b$ in terms of CP-violation contributions to $B_s$ mixing.[5,72] In order to reconcile the D0 result, regarding the same-sign dimuon charge asymmetry, contributions from NP in both $B_d^0 - \overline{B_d^0}$ and $B_s^0 - \overline{B_s^0}$ mixing are needed.[72,73].

The flavor-specific charge asymmetry $a_{s\ell}^q$ is related to the mass and width differences in the $B_q^0 - \overline{B_q^0}$ system as[9,10]

$$a_{s\ell}^q = \mathrm{Im}\frac{\Gamma_{12}^q}{M_{12}^q} = \frac{|\Gamma_{12}^q|}{|M_{12}^q|}\sin\phi_q = \frac{\Delta\Gamma_q}{\Delta M_q}\tan\phi_q. \qquad (21)$$

Here, $M_{12}^q$ and $\Gamma_{12}^q$ are the dispersive and absorptive parts of $B_q^0 - \overline{B_q^0}$ mixing amplitude respectively. In the SM, these asymmetries are suppressed by the small values of $|\Gamma_{12}^q|/|M_{12}^q|$ and $\phi_q$. Hence, prediction of a sizable value of these asymmetries gives a signal for NP. In the SM, since $\tan\phi_d \approx 0.075$, a large enhancement requires fine tunning in $B_d$ sector. Again, an enhancement in $\Delta\Gamma_d$ would imply the NP contribution to the branching ratio of $B_d$ decay modes to be a few percent, which is ruled out by the experiments[63]. On the other hand, in the SM[5,61,62], $\phi_s \approx 0.004$, and the branching ratios of some decay modes, such as $B_s \to \tau^+\tau^-$, have not yet been strongly constrained. NP models[16,74,75] that increase the decay rate of $B_s \to \tau^+\tau^-$ contribute to the absorptive part of $B_s - \overline{B}_s$ mixing and may enhance $\Delta\Gamma_s$. The enhancement in $\Delta\Gamma_s$ corresponds to an enhancement in the branching ratio $B(B_s \to \tau^+\tau^-)$. Hence, the measurement of $B(B_s \to \tau^+\tau^-)$ gives a better understanding of NP involved in $B_s - \overline{B}_s$ mixing. The study of $B_s \to \tau^+\tau^-$ decay[16,74,75] could explain the observed same-sign dimuon charge asymmetry in the B system. This decay should be studied at the LHCb.[76]

The recent LHCb result of $B_s \to J/\psi\phi$ decay gives [67,77,78]

$$\phi_s = -0.001 \pm 0.101 \pm 0.027 \text{ rad},$$

and $\quad \Delta\Gamma_s = 0.116 \pm 0.018 \pm 0.006 \ ps^{-1}. \qquad (22)$

This is the most precise measurement of $\phi_s$ so far and the first direct observation[78] for a non-zero value for $\Delta\Gamma_s$. In the SM[59], $\Delta\Gamma_d/\Gamma_d$ is very small (almost zero):

$$\Delta\Gamma_d/\Gamma_d = 40.9^{+8.9}_{-9.9} \times 10^{-4}. \qquad (23)$$

Any non-zero measurement of this parameter would be a signal for NP,[79] which could explain the anomalous dimuon charge asymmetry observed by D0 collaboration. From the recent study of CPT violation in hadronic and semileptonic B decays, the Belle collaboration[80] has found

$$\Delta\Gamma_d/\Gamma_d = [-1.7 \pm 1.8 \pm 1.1]\times 10^{-2}, \qquad (24)$$

which is an order of magnitude larger than the SM prediction. Recently[81], the mixing phase is found to be:

$$\phi_d = 0.74 \pm 0.03 . \tag{25}$$

The size of NP contribution in the $B_s$ system is constrained by the measurement of $\Delta M_s$ alone but in the $B_d$ system it is different.[5] In $B_d$ system, the mass difference $\Delta M_d$ strongly depends on the Wolfenstein parameters $\bar{\rho}$ and $\bar{\eta}$ but this dependence is very weak for $\Delta M_s$. Further, the constraint on NP in $B_d$ mixing depends on $|V_{ub}|$ and CKM angle $\gamma$. In this paper, we study the $B_q^0 - \overline{B_q^0}$ mixing in a $Z'$ model to explain the large same-sign dimuon charge asymmetry observed by the D0 collaboration. We discuss the NP contributions of $Z'$ to both $M_{12}^q$ and $\Gamma_{12}^q$ considering three different cases that affect the value of the same-sign dimuon charge asymmetry.

## 3. The Model

In extended quark sector model[41–43,82], besides the three standard generations of the quarks, there is an $SU(2)_L$ singlet of charge $-1/3$. This model allows for Z-mediated FCNCs. The up quark sector interaction eigenstates are identified with mass eigenstates but down quark sector interaction eigenstates are related to the mass eigenstates by a 4 × 4 unitary matrix, which is denoted by K. The charged-current interactions are described by

$$L_{\text{int}}^W = \frac{g}{\sqrt{2}} \left( W_\mu^- J^{\mu^+} + W_\mu^+ J^{\mu^-} \right), \tag{26}$$

$$J^{\mu^-} = V_{ij} \bar{u}_{iL} \gamma^\mu d_{jL} . \tag{27}$$

The charged-current mixing matrix V is a 3 × 4 submatrix of K :

$$V_{ij} = K_{ij} \quad \text{for } i = 1,......3, \quad j = 1,......,..4 . \tag{28}$$

Here, V is parametrized by six real angles and three phases, instead of three angles and one phase in the original CKM matrix.

The neutral-current interactions are described by

$$L_{\text{int}}^Z = \frac{g}{\cos \theta_W} Z_\mu \left( J^{\mu 3} - \sin^2 \theta_W J_{em}^\mu \right) , \tag{29}$$

$$J^{\mu 3} = -\frac{1}{2} U_{pq} \bar{d}_{pL} \gamma^\mu d_{qL} + \frac{1}{2} \delta_{ij} \bar{u}_{iL} \gamma^\mu u_{jL} . \tag{30}$$

In neutral-current mixing, the matrix for the down sector is U = V$^\dagger$V. Since in this case V is not unitary, $U \neq 1$. Its non-diagonal elements do not vanish:

$$U_{pq} = - K_{4p}^* K_{4q} \quad \text{for } p \neq q . \tag{31}$$

Since the various $U_{pq}$ are non-vanishing, they allow for FCNCs that would be a signal for NP.

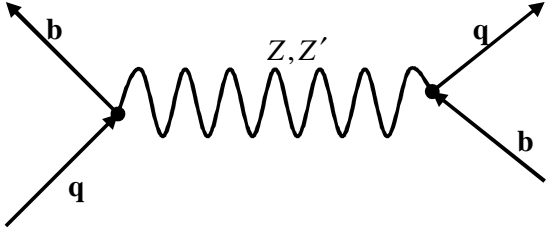

(a)

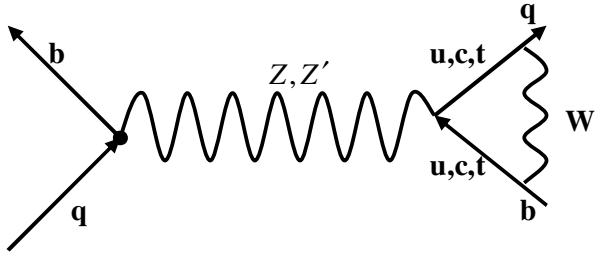

(b)

**Fig. 2:** Feynman diagrams for $B_q^0 - \overline{B_q^0}$ (q = s, d) mixing in the extended quark model, where the blob represents the tree level flavour changing vertex.

Now consider the $B_q^0 - \overline{B_q^0}$ mixing (q = d, s) in the presence of Z-mediated FCNC[41–43,82] at tree level (Fig. 2).[83,84] The Z-mediated FCNC couplings $U_{db}^Z$ and $U_{sb}^Z$, which affect the $B_q^0 - \overline{B_q^0}$ mixing, are constrained from the experimental limit on $B(B \to \ell^+\ell^- X)$.[44–47] The Z-mediated flavour-changing couplings $U_{qb}^Z$ can contribute to $B_q^0 - \overline{B_q^0}$ mixing:[82]

$$M_{12}^Z(B_q) = \frac{\sqrt{2} G_F M_{B_q} \eta_{B_q}}{12} f_{B_q}^2 B_{B_q} (U_{qb}^Z)^2 . \tag{32}$$

Now consider the $B_q^0 - \overline{B_q^0}$ mixing (q = d, s) in the presence of $Z'$-mediated FCNC at tree level. The mixing among particles which have different $Z'$ quantum numbers will induce FCNCs due to $Z'$ exchange.[40,48–50,85–88] For example, if the right-handed ordinary quarks $d_R, s_R$ and $b_R$ have different $Z'$ quantum numbers than exotic quark $h_R$, then the mixing of these ordinary and exotic quarks induces $Z'$-mediated FCNCs. Since the $U_{pq}^Z$ are generated by mixing that breaks weak isospin, they are expected to be at most $O(M_1/M_2)$, where $M_1(M_2)$ is typical light (heavy) fermion mass. On the other hand, the $Z'$-mediated coupling $U_{pq}^{Z'}$ can be generated via mixing of particles with same weak isospin. The new contribution from $Z'$ boson is exactly in the similar manner as in the Z boson (Fig. 2).[83,84] Therefore, the contribution of $Z'$-mediated FCNCs to $B_q^0 - \overline{B_q^0}$ mixing[40] is,

$$M_{12}^{Z'}(B_q) = \frac{\sqrt{2} G_F M_{B_q} \eta_{B_q}}{12} \frac{M_Z^2}{M_{Z'}^2} f_{B_q}^2 B_{B_q} (U_{qb}^{Z'})^2 . \tag{33}$$

Now considering the contributions from $Z'$-mediated FCNC, we can write the new mass matrix element for $B_q^0 - \overline{B_q^0}$ mixing as:[1]

$$M_{12}(B_q) = M_{12}^{SM}(B_q) + M_{12}^{Z'}(B_q). \tag{34}$$

## 4. Evaluation of $B_q^0 - \overline{B_q^0}$ mixing mass differences

The $B_q^0 - \overline{B_q^0}$ (q = s,d) mixing mass differences can be evaluated by substituting equations (4), (33) and (34) in equation (6). Thus, considering the contributions from $Z'$-mediated FCNC, we can write the $B_s^0 - \overline{B_s^0}$ mass difference as:

$$\Delta M_{B_s} = 2 \left[ \frac{G_F^2 M_W^2 M_{B_s} \eta_{B_s}}{12 \pi^2} f_{B_s}^2 B_{B_s} S_0(x_t)(V_{ts}^* V_{tb})^2 + \frac{\sqrt{2} G_F M_{B_s} \eta_{B_s}}{12} \frac{M_Z^2}{M_{Z'}^2} f_{B_s}^2 B_{B_s} (U_{sb}^{Z'})^2 \right]$$

………….(35)

Similarly, the $B_d^0 - \overline{B_d^0}$ mass difference can be written as:

$\Delta M_{B_d}$

$$= 2\left[\frac{G_F^2 M_W^2 M_{B_d} \eta_{B_d}}{12\pi^2} f_{B_d}^2 B_{B_d} S_0(x_t)(V_{td}^* V_{tb})^2 + \frac{\sqrt{2} G_F M_{B_d} \eta_{B_d}}{12} \frac{M_Z^2}{M_{Z'}^2} f_{B_d}^2 B_{B_d} (U_{db}^{Z'})^2\right]$$

……….(36)

The equations (35) and (36) are used in the next section for our calculations.

## 5. Numerical Results and Discussions

We have taken the recent data from:[6] $G_F = (1.16637 \pm 0.00001) \times 10^{-5} \text{GeV}^{-2}$, $M_{B_s} = (5366.0 \pm 0.9)$ MeV, $M_W = (80.399 \pm 0.23)$ GeV, $m_t = 172.0 \pm 0.9 \pm 1.3$ GeV, $M_{B_d} = (5279.5 \pm 0.5)$ MeV, $M_Z = (91.1876 \pm 0.0021)$ GeV. Using the lattice QCD calculations,[89] $f_{B_d}\sqrt{B_{B_d}} = (216 \pm 9 \pm 13)$ MeV, $f_{B_s}\sqrt{B_{B_s}} = (275 \pm 7 \pm 13)$ MeV and assuming $|V_{tb}| = 1$, one finds $|V_{td}| = (8.4 \pm 0.6) \times 10^{-3}$, and $|V_{ts}| = (38.7 \pm 2.1) \times 10^{-3}$. The Inami-Lim function[5] $S_0 = 2.35$, and $\eta_{B_s} = \eta_{B_d} = 0.552$. The value of $|U_{sb}^Z| \cong 10^{-3}$ and $|U_{db}^Z| \cong 10^{-3}$.[41–43] From the study of $B_s^0 - \bar{B}_s^0$ mixing in leptophobic $Z'$ model, they[4] obtained $|U_{sb}^{Z'}| \leq 0.036$ for $M_{Z'} = 700$ GeV and $|U_{sb}^{Z'}| \leq 0.051$ for $M_{Z'} = 1$ TeV. We take $|U_{sb}^{Z'}| \approx 0.03$ and $|U_{db}^{Z'}| \approx 7.8 \times 10^{-3}$ for our calculations. The constraint of $B^+ \to K^+ \tau^+ \tau^-$ from[90] gives the bound $|\Gamma_{12}^{q,NP}/M_{12}^{q,SM}| < 0.3$. However, the presence of $(\bar{s}b)(\bar{\tau}\tau)$ operators[12] can enhance $\Gamma_{12}^s$ to maximum 35 % compared to its SM value. The NP in $\Gamma_{12}^s$ in form of $(\bar{s}b)(\bar{\tau}\tau)$ operators is only able to partly resolve the anomaly in the dimuon charge asymmetry observed recently by the D0 collaboration. In this paper, we use the value of NP effects in $\Gamma_{12}^q$ as:

$$(\Gamma_{12}^q)^{NP} = 0.3 \times (\Gamma_{12}^q)^{SM}. \tag{37}$$

Since the $Z'$ boson has not yet been found, its exact mass is unknown. However, the $Z'$ mass is constrained by direct searches at Fermilab, weak neutral current data and precision studies at LEP and the SLC,[29–34] which give a model-dependent lower bound around 500 GeV. In a study of $B$ meson decays with $Z'$-mediated FCNCs, they[87] study the $Z'$ boson in the mass range of a few hundred GeV to 1 TeV. From the recent CMS collaboration analyses[91] the lower mass limits for the sequential standard model $Z'$ and the superstring inspired $Z'_\psi$ are about 2590 GeV and 2260 GeV respectively at 95% C.L. Dittmar, Nicollerat and Djouadi[92] have studied $Z'$ boson at LHC. They confirm that $Z'$ bosons can be observed in the process $pp \to Z' \to \ell^+ \ell^-$ ($\ell = e, \mu$), up to masses of about 5 TeV. In this paper, for our calculations we take the mass of $Z'$ boson in the range 500 GeV–5 TeV. Using equations (16), (21), (35), (36) and (37), we calculate the same-sign dimuon charge asymmetry for $B_q^0 - \bar{B}_q^0$ system. Here, we consider three different scenarios.

**(a) First scenario (S1):** When the $Z'$ boson contributes only to the off-diagonal element $M_{12}^q$ i.e. $\left(M_{12}^q\right)_{Z'} \neq 0$ and $\left(\Gamma_{12}^q\right)_{Z'} = 0$.

Using the mass of $Z'$ boson, $M_{Z'} = 500$ GeV and all recent data we get

$$A_{s\ell}^b = (-2.8 \pm 0.1) \times 10^{-4}. \tag{38}$$

Again using the mass of $Z'$ boson, $M_{Z'} = 5$ TeV and all recent data we get

$$A_{s\ell}^b = (-2.6 \pm 0.2) \times 10^{-4}. \tag{39}$$

**(b) Second scenario (S2):** When $Z'$ boson contributes only to $\Gamma_{12}^q$ i.e. $\left(M_{12}^q\right)_{Z'} = 0$ and $\left(\Gamma_{12}^q\right)_{Z'} \neq 0$.

$$A_{s\ell}^b = (-3.1 \pm 0.5) \times 10^{-4}. \tag{40}$$

**(c) Third scenario (S3):** In this scenario, $Z'$ boson contributes to both $M_{12}^q$ and $\Gamma_{12}^q$ i.e. $\left(M_{12}^q\right)_{Z'} \neq 0$ and $\left(\Gamma_{12}^q\right)_{Z'} \neq 0$.

Using the mass of $Z'$ boson, $M_{Z'} = 500$ GeV and all recent data we get

$$A_{s\ell}^b = (-3.7 \pm 0.2) \times 10^{-4}. \tag{41}$$

Again using the mass of $Z'$ boson, $M_{Z'} = 5$ TeV and all recent data we get

$$A_{s\ell}^b = (-3.4 \pm 0.3) \times 10^{-4}. \tag{42}$$

From equations (38), (39), (41) and (42), it is clear that depending on the precise value of $M_{Z'}$, the $Z'$-mediated FCNCs gives sizable contributions to the $B_q^0 - \overline{B_q^0}$ system. Our results (equations (38)–(42)) satisfy the bound obtained theoretically on the maximum value of dimuon asymmetry[93] $A_{s\ell}^{b,\text{MAX}} \approx (-5 \pm 1) \times 10^{-3}$. Our estimated same-sign dimuon charge asymmetry for $B_q^0 - \overline{B_q^0}$ system is enhanced from its SM prediction [equation (18)]. Lower the mass of $Z'$ boson, more is the deviation from the SM. Hence, the $B_q^0 - \overline{B_q^0}$ mixing could provide signals for NP beyond the SM.

## 6. Conclusions

The D0 collaboration has measured same-sign dimuon charge asymmetry that differs from the SM prediction by $3.9\sigma$ deviations. At present we do not understand the origin of this discrepancy.[94] In order to reconcile the D0 result, contributions from NP in both $B_d^0 - \overline{B_d^0}$ and $B_s^0 - \overline{B_s^0}$ mixing are needed.[72,73] In this paper, we calculate the same-sign dimuon charge asymmetry for $B_q^0 - \overline{B_q^0}$ system. We find that the same-sign dimuon charge asymmetry is enhanced from its SM prediction due to the effect of $Z'$-mediated FCNCs. Although such a $Z'$ model with our specific assumptions could not reproduce the large same-sign dimuon charge asymmetry for $B_q^0 - \overline{B_q^0}$ system observed recently by the D0 collaboration, it provides signals for NP beyond the SM. In order to reproduce the D0 results in our model, the mass of $Z'$ boson should be as low as ~ 16 GeV for S1 and ~ 19 GeV for S3. Recently[95], it is shown that CP violation in mixing of $B_d^0$ and $B_s^0$ mesons, CP violation in interference of $B_d^0$ decay with and without mixing, CP violation in interference of $B_s^0$ decay with and without mixing, and direct CP violation in semileptonic decays of charged and neutral hadrons containing b or c quarks have contributions to the same-sign dimuon charge asymmetry. The same-sign dimuon charge asymmetry is calculated[95] by considering these additional SM source of dimuon charge asymmetry. It is found that the D0 measurements have still $3.0\sigma$ deviations. Therefore, the latest measurements of the same-sign dimuon charge asymmetry of $B_q^0 - \overline{B_q^0}$ mixing by the D0 collaboration is not fully explained so far. The precise measurements of $a_{s\ell}^s$ and $a_{s\ell}^d$ are necessary to determine whether scenarios[96] with NP in $M_{12}^d$, $M_{12}^s$, $\Gamma_{12}^d$ and/or $\Gamma_{12}^s$ are enough to explain the discrepancy in same-sign dimuon charge asymmetry of $B_q^0 - \overline{B_q^0}$ mixing between the SM prediction and the D0 result or there are other ways of approach.

**Acknowledgments**

We would like to thank Dr. P. K. Behera, IIT, Madras, India, Dr. T. K. Nayak, VECC, Kolkata and Dr. Diptimoy Ghosh, TIFR, India for helpful discussions and suggestions. We thank the reviewer for suggesting valuable improvements of our manuscript.